# Uncertainty propagation and covariance analysis of $^{181}$Ta(n,$\gamma$)$^{182}$Ta nuclear reaction

**Namrata Singh[1], Mahesh Choudhary[1], A. Gandhi[1,5], Aman Sharma[1,6], Mahima Upadhyay[1], Punit Dubey[1], Akash Hingu[2], G. Mishra[3], Sukanya De[3], A. Mitra[3], L. S. Danu[3], Ajay Kumar[3], R. G. Thomas[3], Saurav Sood[4], Sajin Prasad[4], and A. Kumar[1]**

[1]Department of Physics, Banaras Hindu University, Varanasi 221005, India
[2]Department of Physics, The Maharaja Sayajirao University of Baroda, Vadodara 390002, India
[3]Nuclear Physics Division, Bhabha Atomic Research Centre, Mumbai 400085, India
[4]Health Physics Division, Bhabha Atomic Research Centre, Mumbai 400085, India
[5]Horia Hulubei National Institute of Physics and Nuclear Engineering - IFIN-HH, Bucharest 077125, Romania
[6]Nuclear and Chemical Science Division, Lawrence Livermore National Laboratory, Livermore, CA 94551, USA

E-mail: ajaytyagi@bhu.ac.in

April 2024

**Abstract.** The neutron capture cross-section for the $^{181}$Ta(n,$\gamma$)$^{182}$Ta reaction has been experimentally measured at the neutron energies 0.53 and 1.05 MeV using off-line $\gamma$-ray spectrometry. $^{115}$In(n,n'$\gamma$)$^{115m}$In is used as a reference monitor reaction cross-section. The neutron was produced via the $^{7}$Li(p,n)$^{7}$Be reaction. The present study measures the cross-sections with their uncertainties and correlation matrix. The self-attenuation process, $\gamma$-ray correction factor, and low background neutron energy contribution have been calculated. The measured neutron spectrum averaged cross-sections of $^{181}$Ta(n,$\gamma$)$^{182}$Ta are discussed and compared with the existing data from the EXFOR database and also with the ENDF/B-VIII.0, TENDL-2019, JENDL-5, JEFF-3.3 evaluated data libraries.

## 1. Introduction

Neutron capture cross-sections provide important information for reactor design, operation, and safety. They have implications for fuel selection, reactor control strategy, efficiency optimization, safety assessments, and radiation protective measures. In nuclear astrophysics, neutron capture cross-sections are essential for comprehending star nucleosynthesis processes like the s- and r-processes, which produce many heavy elements through fast and slow neutron capture, respectively [1, 2, 3]. Tantalum is a valuable material for manufacturing, so its interaction with neutrons is essential. It is used in reactor vessels, control rods, and other structural parts of nuclear reactors, especially in applications where mechanical strength and resistance to corrosion are essential [4, 5, 6, 7]. Nuclear radiation detectors and measurement instruments use tantalum-based materials. Tantalum offers advantageous characteristics for γ-radiation and high-energy particle detection. Due to their sensitivity and dependability, tantalum-based detectors are used in radiation monitoring systems, medical imaging equipment, and nuclear physics research. After the neutron capture, $^{181}$Ta becomes $^{182}$Ta excited compound nucleus with a positive Q-value of 1815 keV and it decays to $^{182}$W stable nucleus via the β-decay process with a half-life $t_{1/2}$=114.74±0.12 days. During this β-decay process, it produces 67.749 keV, 1121.290 keV, and 1221.395 keV γ-rays with 43%, 35%, and 27% γ-ray intensities for determining the neutron capture cross-section for the $^{181}$Ta nucleus, respectively. $^{182}$Ta can be used as a tracer in geochemistry and environmental science studies. In addition to its use in the nuclear program, the experimental capture cross-sections are crucial for validating various statistical model codes and assessing the impact of different parameter sets on cross-section calculations. The standard monitor reaction cross-section $^{115}$In(n,n'γ)$^{115m}$In from the IRDFF-1.05 data library was used as the reference for the measurement. There is one recent data on the present reaction. Since there is no explanation of the covariance analysis, an experiment has been carefully planned and executed to determine the neutron capture cross-section of tantalum with improved precision and uncertainty propagation. These are important for safety and affordability in nuclear applications. The present results have been discussed, and compared to the cross-sectional data already available in the EXFOR database, and evaluated nuclear data libraries ENDF/B-VIII.0, TENDL-2021, JENDL-5, and JEFF-3.3 [8, 9, 10, 11, 12, 13]. The theoretical calculations of the neutron radiative capture cross-section were also performed with the help of the nuclear reaction code TALYS-1.95 [14].

Table 1: Details about the decay data parameters.

| Reaction | Product | $E_\gamma$ (keV) | $I_\gamma$ (%) | $t_{1/2}$ |
|---|---|---|---|---|
| $^{181}$Ta(n,γ) | $^{182}$Ta | 1121.29 | 35.24±0.08 | 114.74±0.12 d |
| $^{115}$In(n,n'γ) | $^{115}$In$^m$ | 336.24 | 45.90±0.10 | 4.486±0.004 h |

## 2. Experimental Details

This specific experiment conducted at BARC (Mumbai) using the FOTIA facility would depend on the research goals and interests of the scientists involved [15]. It could involve anything from studying nuclear reactions and properties of exotic nuclei to investigating materials for applications in nuclear energy or radiation therapy. In the present experiment conducted at the Bhabha Atomic Research Centre (BARC) using the

Table 2: Details about irradiation, cooling and counting times.

| $<E_n>$ (MeV) | Reaction | $t_{irr}$ (sec) | $t_{cool}$ (sec) | $t_{count}$ (sec) |
|---|---|---|---|---|
| 0.53 | | 84600 | 25762 | 21600 |
| 1.05 | $^{181}$Ta(n,γ)$^{182}$Ta | 87300 | 6026 | 30600 |
| 0.53 | | 84600 | 22052 | 1800 |
| 1.05 | $^{115}$In(n,n'γ)$^{115}$In$^m$ | 87300 | 972604 | 1800 |

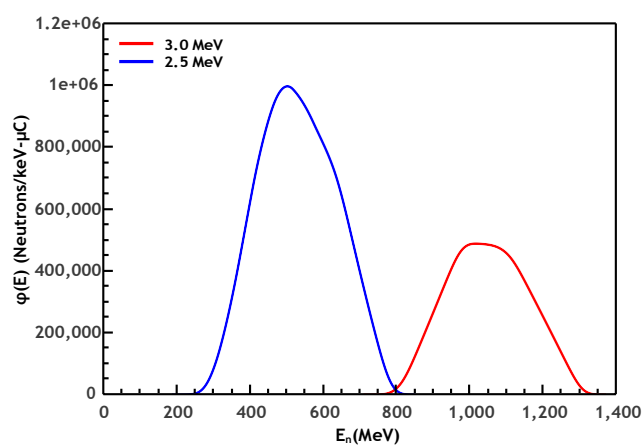

Figure 1: Neutron flux energy spectra at 2.5 and 3.0 MeV proton energies.

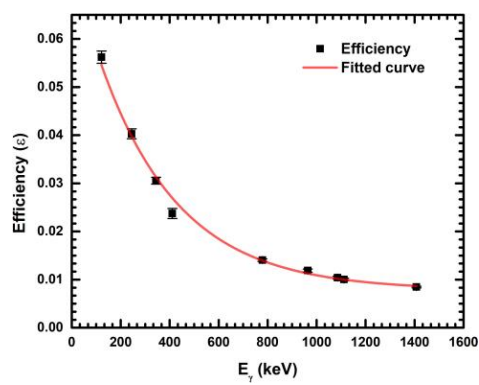

Figure 2: Efficiency calibration curve of the HPGe detector.

Folded Tandem Ion Accelerator (FOTIA) facility, neutrons were produced through the $^7$Li(p,n)$^7$Be nuclear reaction. This reaction involves bombarding lithium-7 ($^7$Li) nuclei with protons (p), resulting in the formation of beryllium-7 ($^7$Be) and emitting neutrons. The $^7$Li(p,n)$^7$Be reaction has a Q-value of -1.644 MeV, and the threshold energy for this reaction is 1.880 MeV. The proton beams with energies of 2.5 MeV and 3.0 MeV were used to bombard a lithium target. By varying the energy of the incident protons, we aimed to produce neutrons with different energies. The code EPEN [16] would have been employed to predict and analyze the neutron energy spectra generated by the $^7$Li(p,n)$^7$Be reaction. EPEN (Energy of Proton Energy of Neutron) is a computational code used for simulating the neutron energy spectra resulting from nuclear reactions induced by proton bombardment. EPEN utilizes theoretical models and experimental data to simulate the interactions between incident protons and target nuclei, predicting the energy distribution of the resulting neutrons. By inputting parameters such as the energy and intensity of the proton beam, properties of the target material, and reaction cross-section data, EPEN calculates the spectrum of emitted neutrons. The spectra of neutron energy, corresponding to two distinct incident proton energies 2.5 and 3.0 MeV, are shown in Figure 1.

In this experiment, we prepared two sets of tantalum to be irradiated with two different energies of neutrons. In addition, we employed natural indium for beam monitoring. To prevent radioactive cross-contamination between the target, monitor, and ambient foils, Al foil was used to wrap the Ta and In foils separately. The samples were irradiated by making a stack of Ta-In of size 10 x 10 mm$^2$ with the neutron beam. Details about decay data parameters are given in Table 1.

Once the irradiation process was finished, the radioactive samples were removed from the irradiation room and subsequently allowed to cool. The neutron irradiation was typically conducted for a duration of approximately 24 hours, interspersed with a cooling period of 2-3 hours before quantifying the gamma activity induced in tantalum. The details about the irradiation, cooling, and counting times are given in Table 2. The irradiated samples of tantalum and indium were mounted on separate perplex plates once they had cooled sufficiently before being transported to the counting room. The induced activity of the product nuclides was assessed utilizing a 185-cc high-purity germanium detector (HPGe) that had been pre-calibrated and shielded with lead. The efficiency calibration of the HPGe detector for various gamma-ray energies has been measured using a standard $^{152}$Eu point source (t$_{1/2}$ = 3.517±0.009 y) with a known activity (A$_0$=6659±82 Bq as on 1 Oct. 1999). The efficiency of the point source was determined by [17, 18]

$$\varepsilon_p = \frac{CK_c}{A_0 I_\gamma e^{-\lambda t} \Delta t} \quad (1)$$

where C is the number of counts, K$_c$ is the coincidence summing factor, A$_0$ is the known activity of the $^{152}$Eu point source, I$_\gamma$ is the intensity of the particular γ-ray energy, the time between date of manufacturing and date of counting is the elapsed time t$_c$ and t is the counting time.

Table 3: Efficiency of the HPGe detector for the point source ($\varepsilon_p$) and sample geometry ($\varepsilon$) with intensities ($I_\gamma$), counts (C) and correction factors ($K_c$).

| $E_\gamma$ (keV) | $I_\gamma$ | Counts (C) | $K_c$ | $\varepsilon_p$ | $\varepsilon$ |
|---|---|---|---|---|---|
| 121.733 | 0.285 ± 0.0016 | 101246 ± 1748 | 1.223 | 0.0584 | 0.0562 ± 0.0013 |
| 244.736 | 0.075 ± 0.0004 | 18143.5 ± 381 | 1.273 | 0.0412 | 0.0403 ± 0.0010 |
| 344.294 | 0.266 ± 0.0020 | 53877.3 ± 916.5 | 1.149 | 0.0314 | 0.0306 ± 0.0007 |
| 411.130 | 0.022 ± 0.00013 | 3017 ± 123.3 | 1.375 | 0.0249 | 0.0237 ± 0.0010 |
| 778.935 | 0.129 ± 0.0008 | 11473.9 ± 135.1 | 1.212 | 0.0145 | 0.0141 ± 0.0002 |
| 867.412 | 0.042 ± 0.0003 | 11600.9 ± 265.3 | 1.094 | 0.0118 | 0.0119 ± 0.0003 |
| 964.077 | 0.145 ± 0.0007 | 8495.2 ± 210.9 | 0.938 | 0.0106 | 0.0104 ± 0.0002 |
| 1085.840 | 0.101 ± 0.0005 | 9747.3 ± 230.6 | 1.031 | 0.0098 | 0.0100 ± 0.0003 |
| 1112.103 | 0.137 ± 0.0008 | 12216.8 ± 158.9 | 1.057 | 0.0083 | 0.0084 ± 0.0002 |
| 1408.030 | 0.209 ± 0.0009 | 125213.2 ± 1496 | 1.039 | 0.0061 | 0.0050 ± 0.00007 |

A correction has been done to adjust for the effects of coincidence summing on gamma lines in the measurement of detector efficiency. The Monte Carlo simulation code EEFTRAN was used to determine the true coincidence summing effect [19, 20]. This simulation needs the exact parameters of the HPGe detector and calibration source. The efficiency values and obtained correction factors ($K_c$) with counts, intensity of gamma lines and gamma energies are given in Table 3. The detector efficiency for the specific γ-ray energy of the resulting nuclides was determined by interpolating the individual efficiencies of the γ-ray energies provided in Table 3 using the fitting function described by

$$\varepsilon(E_\gamma) = \varepsilon_o exp(-E_\gamma/E_0) + \varepsilon_c \qquad (2)$$

where $\varepsilon_o$, $\varepsilon_c$, and $E_0$ were the parameters determined by fitting the above function to the measured efficiencies. The measured efficiencies and the fitted efficiency curve are shown in Figure 2. The parameter values for the fitting are provided in Table 4, along with their corresponding uncertainties and correlation matrix. These values are subsequently used to calculate the covariance matrix for the interpolated detection efficiencies, which can be given in Table 5.

## 3. Theoretical Calculations

TALYS is a computer code system designed to perform nuclear reaction analysis and prediction. It is a Fortran-based code used to compute several physical observables associated with nuclear processes. This nuclear code uses the Hauser-Fesbach statistical model and offers several options for level density and optical model parameters. The fundamental purpose of its development is to simulate nuclear reactions involving particles of $^3$He, alpha, neutrons, photons, protons, deuterons, tritons, and photons within the energy range of 1 keV to 200 MeV, with a focus on target nuclei with masses 12 and greater. When the projectile energy is between 1 keV and several hundreds of MeV, changing the impact energy makes a certain nuclear reaction process more or less important. When discrete-level information is not available, nuclear-level densities are used in statistical models to guess cross sections at activation energies. We use different types of models to figure out the level density in TALYS. These models range from phenomenological analytical statements to tabulated level densities from microscopic models. The TALYS offers six distinct level density models. The ldmodels are as follows: ldmodel-1 for constant temperature and the Fermi gas model [21], ldmodel-2 for the back-shifted Fermi gas model [22], ldmodel-3 for the generalized superfluid model [23, 24], ldmodel-4 for the Goriely table (Skyrme Force) and ldmodel-5 for the Hilaires combinatorial tables (Skyrem force) [25], and ldmodel-6 for the Hilaires combinatorial tables (temperature-dependent HFB, Gogny force) [26]. Among the six level density models, ldmodels 1–3 are phenomenological, whereas ldmodels 4–6 are microscopic.

Table 4: Efficiency ($\varepsilon$) curve fitting parameter values with its uncertainty and correlation matrix.

| Parameters | Value | Uncertainty | Correlation matrix | | |
|---|---|---|---|---|---|
| $\varepsilon_c$ | 0.007 | $4.030 \times 10^{-4}$ | 1.000 | | |
| $\varepsilon_0$ | 0.068 | 0.003 | 0.489 | 1.000 | |
| $E_0$ (keV) | 323.964 | 18.956 | -0.837 | -0.791 | 1.000 |

Table 5: Interpolated detector efficiency of the characteristic $\gamma$-ray of the reaction of the samples with its uncertainty.

| Reaction | $E_\gamma$(keV) | Efficiency | Uncertainty | Covariance matrix | |
|---|---|---|---|---|---|
| $^{181}$Ta(n,$\gamma$)$^{182}$Ta | 1121.29 | 0.00099 | 0.00019 | $3.79003 \times 10^{-8}$ | |
| $^{115}$In(n,n'$\gamma$)$^{115}$In$^m$ | 336.24 | 0.00319 | 0.00067 | $6.79397 \times 10^{-9}$ | $4.60349 \times 10^{-7}$ |

Table 6: Correction factors for γ-ray self-attenuation ($C_\gamma$) and low energy background neutron ($N_{low}$).

| $<E_n>$ | Reaction | $C_\gamma$ | $N_{low}$ |
|---|---|---|---|
| 0.53 | $^{181}$Ta(n,γ)$^{182}$Ta | 1.001 | 0.988 |
| 1.05 | | | 0.909 |
| 0.53 | $^{115}$In(n,n'γ)$^{115}$In$^m$ | 1.025 | 1.003 |
| 1.05 | | | 0.530 |

Table 7: Monitor cross-sections with their uncertainty and correlation matrix.

| $<E_n>$ (MeV) | $<\sigma_m>$ (mb) | Correlation matrix | |
|---|---|---|---|
| 0.53 | 3.899±0.153 | 1.000 | |
| 1.05 | 86.858±2.662 | 0.363 | 1.000 |

Table 8: Measured reaction cross-sections of $^{181}$Ta(n,γ)$^{182}$Ta reaction with their uncertainty and correlation matrix.

| $<E_n>$ (MeV) | $<\sigma_m>$ (mb) | Correlation matrix | |
|---|---|---|---|
| 0.53 | 170.736±9.158 | 1.000 | |
| 1.05 | 118.689±5.676 | 0.571 | 1.000 |

## 4. Data Analysis

*4.1. Cross-section and its uncertainty quantification*

The cross-section for $^{181}$Ta(n,γ)$^{182}$Ta reaction was determined using the following activation formula [27]:

$$<\sigma>_s = <\sigma>_m \times [\eta] \times \frac{A_s I_m \lambda_s a_m N_m f_m \ C_{\gamma(s)} \times N_{low(s)}}{A_m I_s \lambda_m a_s N_s f_s \ C_{\gamma(m)} \times N_{low(m)}} \quad (3)$$

where $\lambda_{s,m}$, $N_{s,m}$, $(A_{s,m})$, $(I_{s,m})$ and $a_{s,m}$ are the decay constant, number of the target atoms, activity of the gamma rays produced, intensity of the gamma rays, and isotopic abundance for sample and monitor respectively. $<\sigma_m>$ is the monitor cross-section. $\eta$ is the ratio of sample and monitor efficiencies. $f_s$ and $f_m$ is the timing factor, which is calculated by:

$$f_s = [1 - exp(-\lambda_s t_{irr})] \times exp(-\lambda_s t_{cool}) \times [1 - exp(-\lambda_s t_{count})] \quad (4)$$

$$f_m = [1 - exp(-\lambda_m t_{irr})] \times exp(-\lambda_m t_{cool}) \times [1 - exp(-\lambda_m t_{count})] \quad (5)$$

where $t_{irr}$, $t_{cool}$ and $t_{count}$ are the irradiation, cooling and counting times.

In the cross-section calculation, the correction factors for the γ-ray self-attenuation $C_γ$ and low-energy background neutron contribution $N_{low}$ have also been included. The following is a discussion of correction factors caused by $C_γ$ and $N_{low}$:

*4.1.1. Background neutron contribution at low energy $N_{low}$* The lower energy background neutrons produced by the $^7$Li(p,n$_1$)$^7$Be reaction contribute to the neutrons produced by the $^7$Li(p,n$_0$)$^7$Be reaction because the incident proton energies in this experiment above the energy threshold of the first excited level of $^7$Be [28]. The (n,γ) reaction cross-section is very sensitive to the neutron energy, so proper calculation regarding it requires the elimination of this (p,n$_1$) neutron background. Thus, we have taken into account the low-energy adjustment and computed by applying the following formula.

$$N_{low} = 1 - \frac{\int \phi_1(E)(\sigma_x(E))dE}{\int \phi(E)(\sigma_x(E))dE} \qquad (6)$$

where, $\phi_1(E)$ is the neutron flux for (p,n$_1$) energy spectrum, $\sigma_x(E)$ is the $^{121}$Sb(n,γ)$^{122}$Sb cross-section obtained from the ENDF/B-VIII.0 and $^{115}$In(n,n'γ)$^{115}$In$^m$ cross-section obtained from the IRDFF-1.05 data library, $\phi(E) = \phi_0(E)+\phi_1(E)$ is the total neutron energy flux taken from the EPEN code. XMuDat 1.0.1 was used to determine the mass attenuation coefficient ($μ_m$) [29].

*4.1.2. Self-attenuation factor for γ-ray $C_γ$* The correction factor for the self-attenuation effect caused by the interaction of -ray inside the samples is investigated in γ-spectrometric analysis [30, 31, 32]. The formula was applied to figure out the correction factor for self-attenuation of γ-ray flux as it traverses a sample with a specified thickness (d), density (ρ), and mass attenuation coefficient ($μ_m$):

$$C_γ = \frac{μ_m ρ d}{1 - \exp(-μ_m ρ d)} \qquad (7)$$

Table 6 provides the values of the low energy correction factor $N_{low}$ and the γ-ray self-attenuation factor $C_γ$.

## 5. Results and Discussion

The $^{181}$Ta(n,γ)$^{182}$Ta reaction cross-sections are measured at neutron energies 0.53, and 1.05 MeV with their uncertainties, covariance, and correlation matrix are given in Table 8 and the data values with their associated uncertainties are plotted in Figure 3 along with the existing cross-sections data obtained from the EXFOR database. Also, we have plotted the ENDF/B-VIII.0, TENDL-2021, JENDL-5, and JEFF-3.3 evaluated nuclear data and theoretically calculated results obtained from TALYS-1.95.

The measured data were found to be in good agreement with the evaluated data, theoretical model predictions, and existing data in the literature. From Figure 3, it can be observed that the present experimental results follow the existing cross-section data trend, i.e. decreasing of capture cross-section with increasing neutron energy. The cross-section data of Poenitz [33] underestimate all other existing data reported by Mcdermott et al. [34], Brzosko et al. [35], Lindner et al. [36], Miskel et al. [37], Bokhovko et al. [38], Kononov et el. [39], Haishan et al. [40] and the evaluated data, theoretical data and also the present result above 0.4 MeV neutron energy.

Table 9: Fractional uncertainties (%) of different parameters for all the two neutron energies.

| Parameters | $<E_n>=0.53$ MeV | $<E_n>=1.05$ MeV |
|---|---|---|
| $A_s$ | 1.25 | 1.26 |
| $A_m$ | 1.79 | 1.09 |
| $I_s$ | 0.227 | 0.227 |
| $I_m$ | 0.218 | 0.218 |
| $N_s$ | 0.174 | 0.193 |
| $N_m$ | 0.093 | 0.082 |
| $a_m$ | 0.052 | 0.052 |
| $\eta$ | 1.083 | 1.083 |
| $f_s$ | 0.0011 | 0.0011 |
| $f_m$ | 0.0892 | 0.0892 |
| $\sigma_m$ | 3.9111 | 3.0647 |
| Total error (%) | 5.31 | 4.78 |

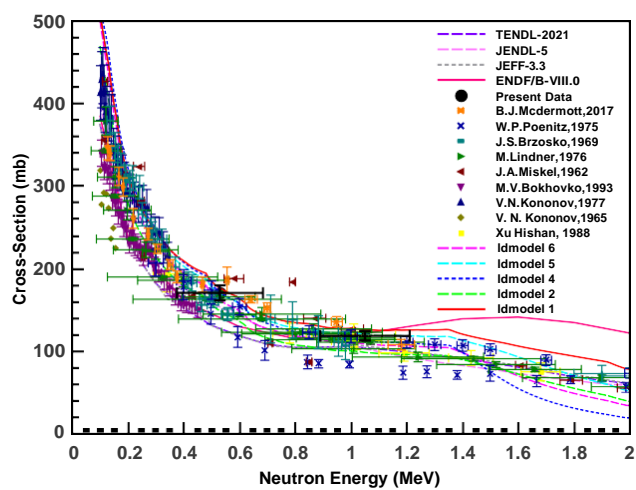

Figure 3: Cross-section of $^{181}$Ta(n,γ)$^{182}$Ta reaction obtained in the present experiment and compared with the evaluated data, existing experimental cross-section data, and the theoretically calculated data from TALYS-1.95.

## 6. Conclusion

Neutron capture cross-section for $^{181}$Ta(n,γ)$^{182}$Ta reaction is carried out at 0.53 and 1.05 MeV neutron energies, relative to the $^{115}$In(n,n'γ)$^{115m}$In standard monitor cross-section followed by off-line γ-ray spectrometry. Neutrons are generated using $^7$Li(p,n) reaction at the FOTIA Facility, Mumbai. We have incorporated important corrections in our analysis such as γ-ray factor and low-energy background neutrons to improve the accuracy. Covariance analysis is used to evaluate the uncertainties associated with the measured reaction cross-sections. We have compared the present data with existing experimental data, ENDF/B-VIII.0, TENDL-2019, JENDL-5, JEFF-3.3 evaluated data and the theoretical data. We have performed for the first time the covariance analysis for $^{181}$Ta(n,γ)$^{182}$Ta nuclear reaction at 0.53 and 1.05 MeV neutron energies. The uncertainties in the measured cross-sections are found to be in the range of 4-6 %. The measured results with detailed uncertainties and covariance are important for the verification of nuclear reaction codes and in the application of nuclear reactors.

**Acknowledgements**

The author (Namrata Singh) is thankful for the financial support of the UGC Non-NET Fellowship (Sanction Letter No.: R/Dev/Sch/UGC Non-NET Fello./2022-23/60710). We acknowledge one of the authors (A. Kumar) for the financial support from IUAC-UGC, Government of India (Sanction No. IUAC/XIII.7/UFR-71353). We would like to extend our heartfelt thanks to the staff of the FOTIA facility for their invaluable support in operating and assisting with the use of the accelerator.

**References**

[1] Gandhi A, Sharma A, Pachuau R, Singh N, Patil P N, Mehta M, Danu L S, Suryanarayana S V, Nayak B K, Lalremruata B and Kumar A 2021 *The European Physical Journal Plus* **136** 819.
[2] Kumar A, Balasubramaniam M, Chakraborty A, Crider B P, Hicks S F, Karthikraj C, Kersting L J, Luke C J, Mcdonough P J, McEllistrem M T and Peters E E 2014 *Journal of Radioanalytical and Nuclear Chemistry* **302** 1043-1047.
[3] Vanhoy J R, Hicks S F, Chakraborty A, Champine B R, Combs B M, Crider B P, Kersting L J, Kumar A, Lueck C J, Liu S H and McDonough P J 2015 *Nuclear Physics A* **939** 121-140.
[4] Shibata K 2016 *Journal of Nuclear Science and Technology* **53** 957-967.
[5] Gritzay O and Libman V 2009, In Reactor Dosimetry State Of The Art 2008, 549-556.
*[6]* Nakamura S, Shibahara Y, Endo S and Kimura A 2021 *Journal of Nuclear Science and Technology* **58** 1061-1070.
[7] Singh N, Gandhi A, Sharma A, Choudhary M and Kumar A 2022 *Indian Journal of Pure and Applied Physics* **58** 314-318.
[8] IAEA-EXFOR Experimental Nuclear Reaction Database. https://www-nds.iaea.org/exfor. Retrieved June 2023.
[9] Otuka N, Dupont E, Semkova V, Pritychenko B, Blokhin A I, Aikawa M, Babykina S et al. 2014 *Nucl. Data Sheets* **120** 272.
[10] Brown D A, Chadwick M B, Capote R, Kahler A C, Trkov A, Herman M W, Sonzogni A A, Danon Y, Carlson A D, Dunn M and Smith D L 2018 *Nuclear Data Sheets* **148** 1-142.
[11] Koning A J, Rochman D, Sublet J C, Dzysiuk N, Fleming M and Van der Marck S 2019 *Nuclear Data Sheets* **155** 1-55.
[12] Liem P H and Hartanto D 2024 *Nuclear Engineering and Design* **418** 112899.
[13] Plompen A J, Cabellos O, De Saint Jean C, Fleming M, Algora A, Angelone M, Archier P, Bauge E, Bersillon O, Blokhin A and Cantargi F 2020 *European Physical Journal A* **56** 1-108.

[14] Koning A J, Hilaire S and Duijvestijn M C, TALYS-1.0, Proceedings of the International Conference on Nuclear Data for Science and Technology, April 22-27, 2007, Nice, France, editors O. Bersillon, F. Gunsing, E. Bauge, R. Jacqmin, and S. Leray, EDP Sciences, 211 (2008).

[15] Singh P, Gupta S K, Kansara M J, Agarwal A, Santra S, Kumar R, Basu A, Sapna P, Sarode S P, Subrahmanyam N B V and Bhatt J P 2002 *Pramana* **59** 739-744.

[16] Pachuau R, Lalremruata B, Otuka N, Hlondo L R, Punte L R M, and Thanga H H 2017 *Nuclear Science and Engineering* **187** 70-80.

[17] Singh N, Choudhary M, Gandhi A, Sharma A, Upadhyay M, Dubey P, Pachuau R, Dasgupta S, Datta J and Kumar A 2024 *European Physical Journal A* **60** 24.

[18] Choudhary M, Sharma A, Gandhi A, Singh N, Dubey P, Upadhyay M, Mishra U, Dubey N K, Dasgupta S, Datta J and Katovsky K 2022 *Journal of Physics G: Nuclear and Particle Physics* **50** 015103.

**[19]** Ramebäck H, Jonsson S, Allard S, Ekberg C and Vidmar T 2015. *J. Radio. Nucl. Chem.* **304** 467-471.

[20] Vidmar T, Kanisch G and Vidmar G 2011 *App. Rad. Iso.* **69** 908-911.

[21] Gilbert A, Cameron A G W 1965 *Can. J. Phys.* **43** 1446.

[22] Dilg W, Schantl W, Vonach H, Uhl M 1973 *Nucl. Phys. A* **217** 269– 411.

[23] Ignatius A V, Istekov K K and Smirenkin G N 1976 *Sov. J. Nucl. Phys.* **29** 450.

[24] Ignatius A V, Weil J L, Raman S and Kahane S 1993 *Phys. Rev. C* **47** 1504.

[25] Goriely S, Hilaire S and Koning A J 2008 *Phys. Rev. C* **78** 064307.

[26] Hilaire S, Girod M, Goriely S and Koning A J 2012 *Phys. Rev. C* **86** 064317.

[27] Choudhary M, Gandhi A, Sharma A, Singh N, Dubey P, Upadhyay M, Dasgupta S, Datta J and Kumar A 2022 *European Physical Journal A* **58** 95.

[28] Gandhi A, Sharma A, Pachuau R, Lalremruata B, Mehta M, Patil P N, Suryanarayana S V, Danu L S, Nayak B K and Kumar A 2021 *European Physical Journal A* **57** 1.

[29] Nowotny R, IAEA Report No. IAEA-NDS 1998 **195**.

[30] Robu E and Giovani C 2009 *Roma. Repo. Phys.* **61** 295-300.

[31] Jackman K R, Ph.D. dissertation submitted to the University of Texas at Austin, August (2007).

[32] Millsap D W and Landsberger S 2015 *App. Rad. Iso.* **97** 21-23.

[33] Poenitz W P, Argonne National Laboratory Reports, No.15 (1975).

[34] McDermott B J, Blain E, Daskalakis A, Thompson N, Youmans A, Choun H J, Steinberger W, Danon Y, Barry D P, Block R C and Epping B E 2017 *Physical Review C* **96** 014607.

[35] Brzosko J S, Gierlik E, Soltan A, Szeflinski Z, Wilhelmi Z 1971 *Acta Physica Polonica* 489.

[36] Lindner M, Nagle R J and Landrum J H 1976 *Nuclear Science and Engineering* **59** 381-394.

[37] Miskel J A, Marsh K V, Lindner M and Nagle R J 1962 *Physical Review* **128** 2717.

[38] Bokhovko M V, Voevodskiy A A, Kononov V N, Poletaev E D and Timokhov V M 1993 Fiz.-Energ Institut, Obninsk Reports, No.2169.

[39] Kononov V N, Stavisskii Yu Ya, Kolesov V E, Dovbenko A G, Nesterenko S and Moroka V I 1965 Fiz.-Energ Institut, Obninsk Reports, No.29.

[40] Haishan Xu, Zhengyu Xiang, Yunshan Mu, Yaoshun Chen and Jinrong Liu 1986 *Chinese J. of Nuclear Techniques* **9** 5.